# Decoding Decision Correctness from EEG Under High Cognitive Workload in Virtual Reality: Implications for Collaborative Brain-Computer Interface Teams

Christopher Baker (corresponding author)
c.baker@qub.ac.uk
School of Electronics, Electrical Engineering and Computer Science
Queen's University Belfast, Belfast, United Kingdom

Stephen Hinton
s.f.hinton@ljmu.ac.uk
School of Psychology
Liverpool John Moores University, Liverpool, United Kingdom

Tom Reed
treed@mail.dstl.gov.uk
Defence Science Technology Laboratory, Salisbury, United Kingdom

Stephen Fairclough
s.fairclough@ljmu.ac.uk
School of Psychology
Liverpool John Moores University, Liverpool, United Kingdom

## Abstract

Collaborative Brain-Computer Interfaces (cBCIs) offer a promising mechanism to augment team decision-making, but existing approaches rely exclusively on evidence available only after a decision has been made and reported, such as reaction time or stated confidence. This limits their use to explaining or discounting a decision after the fact, rather than informing a team's response before it is finalised. We tested whether spatial-covariance EEG features could instead provide a genuinely pre-emptive signal of an operator's decision correctness, available within the response window itself, and whether such a signal depends on cognitive workload. Using a continuous virtual reality target-detection task, participants (N = 23) completed a within-subject workload manipulation (High vs. Low). A Riemannian tangent-space classifier, trained on a strictly motor-safe EEG window preceding each response, decoded Correct from Incorrect decisions with above-chance accuracy under High Workload (61.4%, p = .0014, N = 23) but not reliably under Low Workload (N = 17), where near-ceiling behavioural accuracy left too few error trials for valid classification. Ablation analysis showed the signal was carried predominantly by posterior occipito-parietal channels, consistent with a perceptual rather than motor origin. At the team level, weighting votes by this pre-emptive neural signal, available before a response is committed, produced substantial accuracy gains on contested (evenly-split) trials under High Workload (57% to 88% as team size increased from 2 to 16), but was actively detrimental under Low Workload. Critically, this advantage held even against post-hoc behavioural signals: confidence was the strongest single team-level signal overall, but by definition cannot inform a decision still in progress, whereas the neural signal can. These findings indicate that EEG-based decision-reliability signals are not a general-purpose team augmentation tool, but a workload-conditional one, with clear implications for when and how cBCI systems should be deployed in operational teams.

## Introduction

Modern operational environments increasingly rely on teams of human operators to make rapid perceptual decisions under demanding conditions, monitoring surveillance feeds, screening medical images, or detecting threats in degraded visibility (Cabrera et al., 2023; Canonico et al., 2019; Dehais et al., 2022; Goldfarb and Lindsay, 2022; Koopman and Zammit-Mangion, 2024). The statistical advantage of aggregating such teams, often termed the "wisdom of crowds", is fundamentally predicated on the independence and reliability of individual judgments (Galton, 1907; Hamada et al., 2020; Kerr and Tindale, 2004). However, cognitive workload systematically degrades individual perceptual reliability (Kosch et al., 2023; Plainis and Murray, 2002; Wickens, 2008), and conventional team aggregation methods, whether simple majority vote or approaches weighted by reaction time or past performance, implicitly assume that an operator's explicit behavioural output remains a trustworthy proxy for their underlying perceptual state (Arshad et al., 2015; Bhattacharyya et al., 2019; Cinel et al., 2022; Poli et al., 2014; Woolley et al., 2010). Under sustained high workload, this assumption may not hold.

Contemporary models of metacognition suggest that the sensory evidence contributing to a perceptual decision can be computationally dissociated from the process governing its behavioural report (Fleming and Lau, 2014; Wokke et al., 2020, 2017). Electrophysiological work demonstrates that neural markers of decision correctness, and even of error commission, can be detected independently of, and sometimes prior to, an operator's own explicit awareness or report (Di Gregorio et al., 2020; Murayama et al., 2016; Sadras et al., 2023; Schnuerch et al., 2024; Yeung and Summerfield, 2012). This raises the possibility that an operator's true decision reliability is only partially observable through the behavioural channels, such as reaction time or stated confidence, that current team aggregation methods rely on. Critically, these behavioural channels are also only available after a decision has been reported, meaning conventional aggregation can, at best, discount an unreliable vote after the fact. A signal available earlier, within the decision process itself, would instead allow team aggregation to weight a vote appropriately at the moment it is cast.

Collaborative Brain-Computer Interfaces (cBCIs) have been proposed as a mechanism to give team aggregation access to this otherwise unobservable channel, using neural signals to weight or inform collective decisions directly (Bhattacharyya et al., 2021; Salvatore et al., 2022; Valeriani et al., 2022, 2019, 2017). In continuous, self-paced, ecologically valid environments such as virtual reality, where operators visually acquire a target before any formal decision cue appears, traditional time-domain analyses are vulnerable to substantial latency jitter between stimulus onset and neural response (Hassan et al., 2024; Quitadamo et al., 2017; Zhou et al., 2022). Spatial covariance approaches using Riemannian geometry offer a robust alternative, capturing the relational structure of oscillatory brain activity across the full electrode array rather than depending on precise temporal alignment (Barachant et al., 2012; Congedo et al., 2017; Yger et al., 2018).

The present study applies this approach to a specific and operationally relevant decoding target: not what an operator saw, but whether their judgment about it was correct. This distinction matters because team aggregation ultimately depends on knowing which operators to trust on a given trial, not on independently verifying the world state each operator observed. Using a continuous virtual reality target-detection task under a within-subject manipulation of cognitive workload, we test, first, whether a Riemannian spatial-covariance classifier, trained on a strictly motor-safe EEG window, can decode decision correctness; second, whether this signal is conditional on workload, and what its likely neural origin is; and third, whether it provides real, translatable value when aggregated across simulated teams, evaluated specifically against, not merely alongside, the behavioural signals (confidence, reaction time) that conventional aggregation already has access to. In doing so, we establish the specific operational conditions under which such a signal should be deployed.

## Methods

### Participants

Twenty-three participants (Mean age ± SD = 24.8 ± 6.0; 10 Female) took part in the study, following exclusion of participants during preliminary quality control for persistent LabStreamingLayer synchronisation failures or excessive baseline EEG artefacts. All participants reported normal or corrected-to-normal vision, no history of neurological disorders, and no susceptibility to VR-induced motion sickness. Written informed consent was obtained from all individuals. The experimental protocol received favourable opinion from the UK Ministry of Defence Research Ethics Committee (MoDREC, Reference: RQ0000037929), and all procedures were conducted in accordance with the Declaration of Helsinki.

### Study Design and VR Target-Detection Task

Participants completed a within-subject repeated-measures design within a continuous Virtual Reality (VR) target-detection task. Participants viewed a high-fidelity 3D simulation rendered in Unreal Engine 5 via a Varjo Aero Head-Mounted Display (Varjo Technologies, Helsinki, Finland), depicting the viewpoint of a drone traversing a landscape. Participants discriminated between continuously appearing 3D models, categorised as Non-Targets and Targets (Fig. 1A, 1B). A targeting reticle locked onto each stimulus for 2500 ms, prompting a physical joystick button press before the reticle disappeared (Fig. 2).

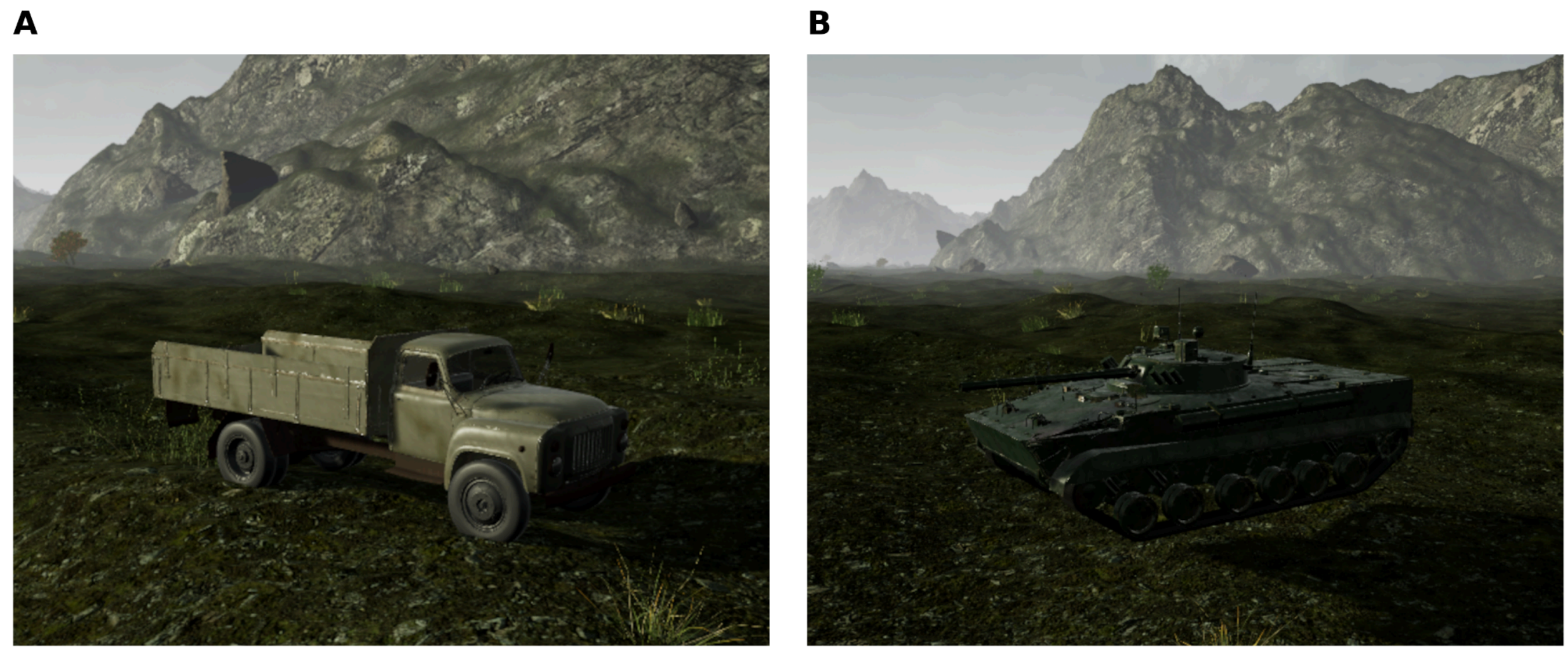

*Figure 1. Virtual Reality target detection task stimuli. (A) Example of a Non-Target stimulus. (B) Example of a Target stimulus.*

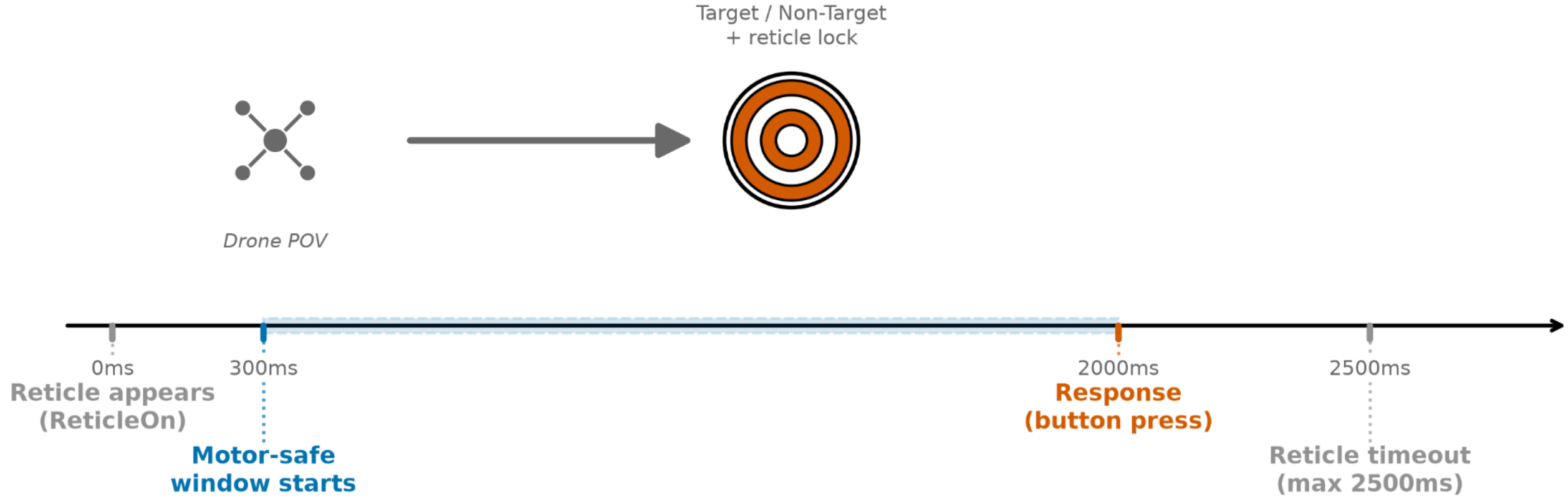

*Figure 2. Trial timeline. A targeting reticle locks onto a Target or Non-Target stimulus at 0 ms and remains locked for a maximum of 2500 ms, within which the participant executes a joystick button press. All EEG features are extracted from a motor-safe window beginning 300 ms post-reticle and ending 100 ms before each trial's own response time.*

Cognitive workload was manipulated within-subject across sessions by simulating a Degraded Visual Environment: High Workload sessions reduced ambient lighting by 50% and lowered the solar angle relative to Low Workload sessions, which used standard daylight conditions (Fig. 3A, 3B).

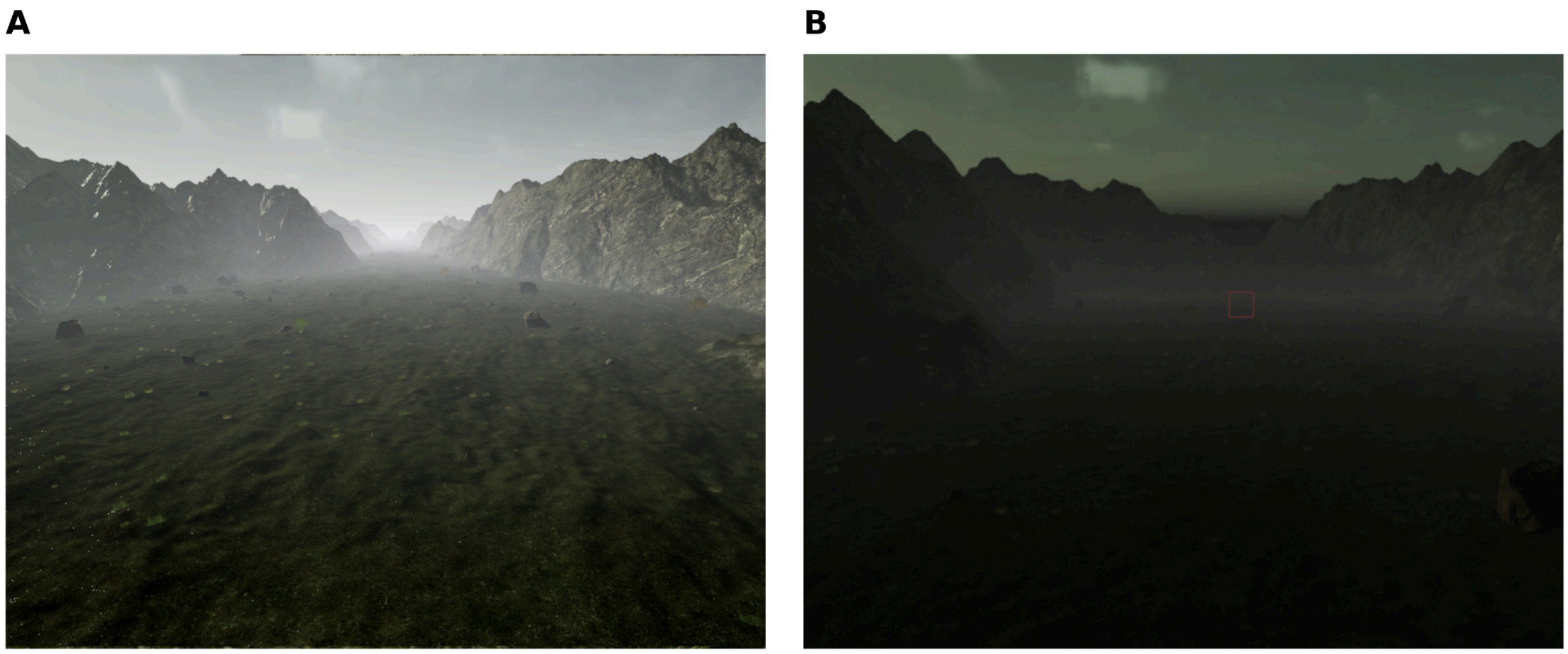

*Figure 3. Environmental Workload Manipulations. (A) Low Workload standard daylight condition. (B) High Workload simulating a Degraded Visual Environment (DVE) via a 50% reduction in lighting.*

## Data Acquisition and EEG Signal Processing

Continuous EEG data were recorded at 500 Hz using a 32-channel LiveAmp system (Brain Products GmbH). Behavioural data and VR events were synchronised via LabStreamingLayer (LSL) to ensure sub-millisecond precision (Dirican and Göktürk, 2011; Kothe et al., 2024). EEG data were processed offline using MNE-Python (Gramfort et al., 2013). Following band-pass (0.1 to 30 Hz) and notch (50 Hz) filtering, Independent Component Analysis (ICA) was computed and components reflecting ocular artefacts were manually rejected(“Faster Independent Component Analysis by Preconditioning With Hessian Approximations | IEEE Journals & Magazine | IEEE Xplore,” n.d.). Data were segmented into epochs (-200 ms to +800 ms relative to reticle onset).

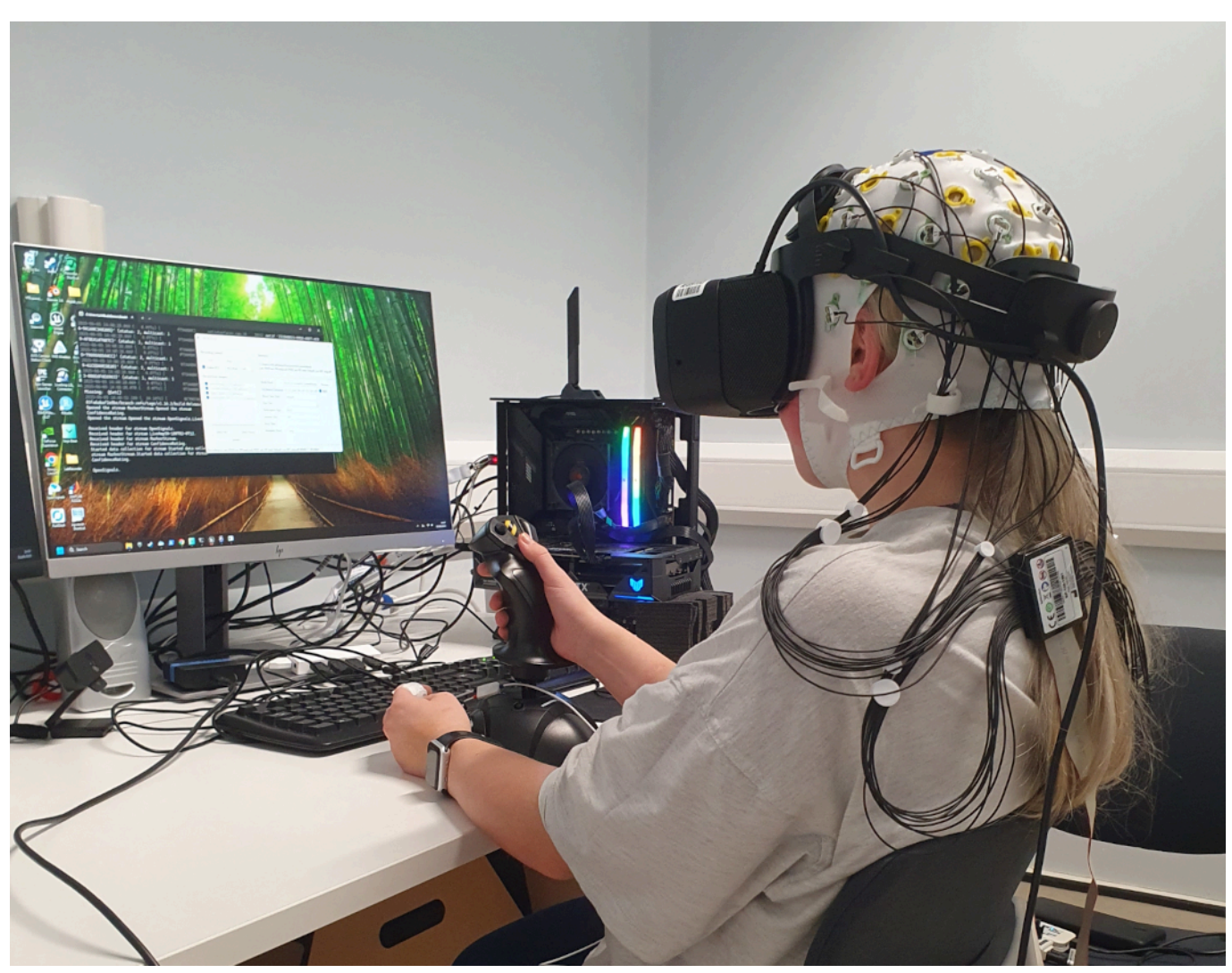

*Figure 4. Experimental setup demonstrating a participant wearing the active EEG cap and VR Head-Mounted Display.*

### Motor-Safe Feature Extraction and Classification

To decode decision correctness independently of motor-execution artefacts, features were extracted from a strictly motor-safe window: from 300 ms post-reticle-onset to 100 ms before each trial's own response time, with a minimum required window duration of 150 ms; trials not meeting this criterion were excluded (Fig. 5). Within this window, a spatial covariance matrix was estimated per trial using Ledoit-Wolf shrinkage (Ledoit and Wolf, 2004) and projected into a Euclidean tangent space via the Riemannian mean (Barachant et al., 2012; Congedo et al., 2017; Yger et al., 2018), using PyRiemann. Features were standardised and classified using logistic regression with balanced class weighting (C = 0.1)(Pedregosa et al., 2011). Model validation used leave-one-subject-out (LOSO) cross-validation, separately for High and Low Workload.

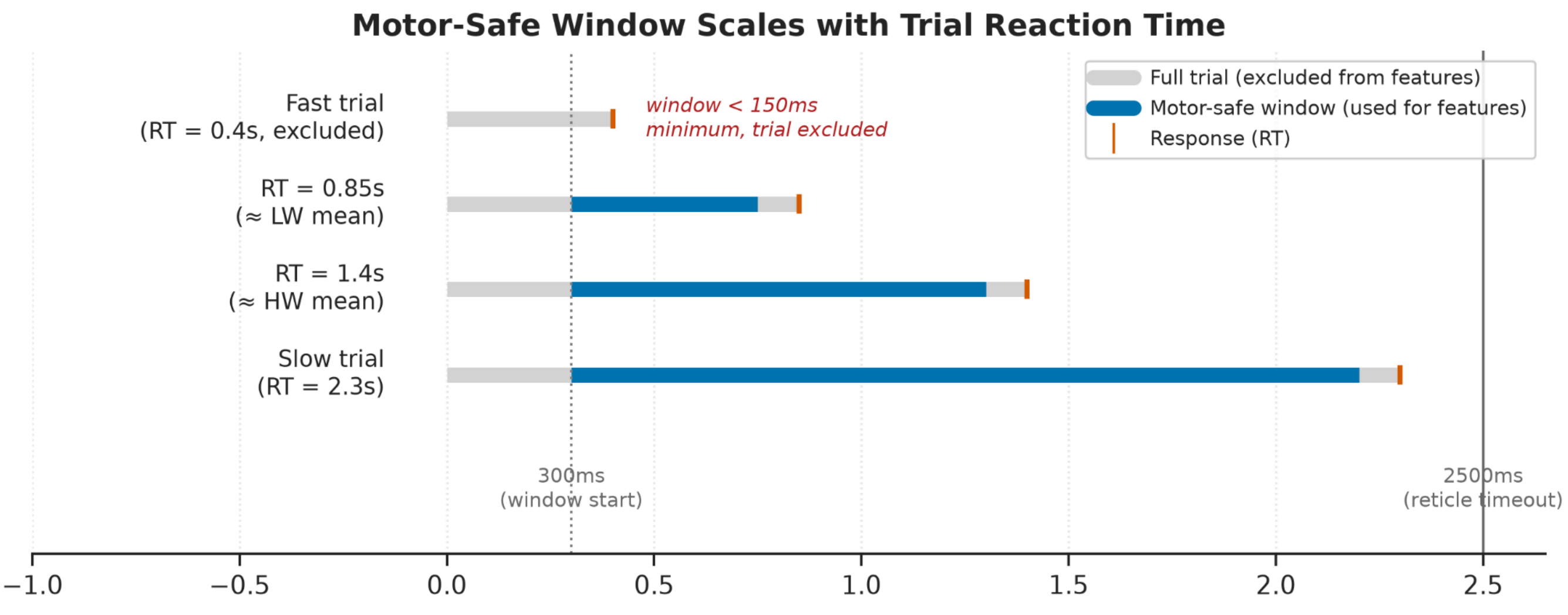

*Figure 5. The motor-safe window scales with each trial's own reaction time. Trials whose window would fall below a 150 ms minimum duration are excluded from feature extraction (top row). Illustrative reaction times shown correspond approximately to the session means for Low Workload (≈0.85 s) and High Workload (≈1.4 s), and to the reticle's 2500 ms hard timeout.*

### Team Simulation and Aggregation

Team-level performance was evaluated via bootstrap resampling (500 draws per team size and session)(Harris et al., 2020; McKinney, 2010) across team sizes of 2, 4, 8, and 16. For each simulated team, votes from individual members' recorded response directions were aggregated by unweighted majority vote, and separately by evidence-weighted vote using: (a) the pre-emptive neural signal (each member's out-of-fold classifier probability, transformed to a signed evidence value), (b) post-hoc stated confidence, and (c) post-hoc response time. To isolate trials where aggregation weighting could plausibly influence the outcome, a secondary analysis restricted to true-tie trials, defined as trials on which a simulated team's raw, unweighted vote was split within one vote of even.

### Statistical Analysis

Paired t-tests assessed the workload manipulation's behavioural effect (accuracy, reaction time) and the Target-detection/Non-Target-rejection asymmetry. One-sample t-tests against chance (50%) assessed classifier accuracy across LOSO folds. Cluster-based permutation testing (5000 permutations) assessed reticle-locked EEG responses against a pre-stimulus baseline(Maris and Oostenveld, 2007). Channel-level contribution to classifier accuracy was assessed via ablation: each channel's covariance row and column were zeroed in turn, and the resulting drop in LOSO accuracy relative to the full-channel baseline was taken as that channel's importance.

## Results

### Workload Manipulation and Behavioural Asymmetry

The workload manipulation produced a robust behavioural effect. Accuracy was significantly lower under High Workload than Low Workload (77.55% vs. 93.84%), and response time was significantly slower (1.40s vs. 0.81s; both $p < .0001$; Table 1), confirming the manipulation successfully degraded perceptual performance. Independently of workload, Target detection was significantly harder than Non-Target rejection ($p = .007$; Table 2), an asymmetry that recurs later in the analysis of trial difficulty.

| Condition | Mean Accuracy | Mean RT |
|---|---|---|
| High Workload | 0.78 | 1.40 |
| Low Workload | 0.94 | 0.81 |

| Metric | T | dof | alternative | p_val | CI95 | cohen_d | power | BF10 |
|---|---|---|---|---|---|---|---|---|
| Accuracy (High vs Low) | -8.8373 | 21 | two-sided | > 0.001 | [-0.2 -0.12] | 1.5881 | 1 | 8.16E+05 |
| RT (High vs Low) | 13.3937 | 21 | two-sided | > 0.001 | [0.5 0.68] | 1.8959 | 1 | 8.78E+08 |

*Table 1. Workload manipulation check. Mean accuracy and response time under High and Low Workload, with paired t-tests comparing the two conditions.*

| Trial Type | Mean Accuracy |
|---|---|
| Target Detection (Hit Rate) | 0.81 |
| Non-Target Rejection (Correct Rejection Rate) | 0.90 |

| T | dof | alternative | p_val | CI95 | cohen_d | power | BF10 |
|---|---|---|---|---|---|---|---|
| -2.6961 | 22 | two-sided | 0.0132 | [-0.16 -0.02] | 0.625 | 0.8172 | 3.894 |

*Table 2. Target detection versus Non-Target rejection accuracy, paired t-test across participants.*

### The Reticle Elicits a Genuine, Motor-Safe Neural Response

Before testing whether EEG distinguishes correct from incorrect decisions, we confirmed that the pipeline detects a genuine evoked response to the reticle at all, independent of decision correctness (Fig. 3, Fig. 5). Using a reaction-time-masked analysis, in which each timepoint's average included only trials whose response had not yet occurred, cluster-based permutation testing identified significant reticle-locked activity in both High Workload (two clusters, $p = .013$ and $p = .027$, spanning approximately 0.31 to 0.80 s post-reticle) and Low Workload (one cluster, $p = .028$, spanning approximately 0.28 to 0.80 s), confirming the pipeline reliably detects real signal within the motor-safe window used for all subsequent analyses.

## EEG Decodes Decision Correctness Under High, but Not Low, Workload

A Riemannian tangent-space classifier, trained on motor-safe spatial covariance features and validated via leave-one-subject-out cross-validation, decoded Correct from Incorrect decisions with above-chance accuracy under High Workload (61.4%, TPR 64.8%, TNR 46.3%, $t(22) = 3.65$, $p = .0014$, N = 23; Table 3). Under Low Workload, raw classifier accuracy was numerically higher (79.4%) but not a comparable finding: true negative rate collapsed to 12.3%, indicating the classifier almost never correctly identified an incorrect trial (Table 3). This is attributable to the behavioural ceiling effect established above (93.84% Low Workload accuracy), which left too few genuine error trials for the classifier to learn a reliable decision boundary. All subsequent analyses of classifier structure and team-level value therefore focus on High Workload; the Low Workload classifier is reported but not mechanistically interpreted.

| Condition | Mean_Accuracy | Mean_TPR_correct | Mean_TNR_incorrect | N_Subjects | t_vs_chance | p_vs_chance |
|---|---|---|---|---|---|---|
| High Workload | 0.61 | 0.65 | 0.46 | 23.00 | 3.65 | 0.0014 |
| Low Workload | 0.79 | 0.86 | 0.12 | 17.00 | 8.66 | 0.0000002 |

*Table 3. Leave-one-subject-out classifier performance decoding Correct versus Incorrect decisions, by workload condition. TPR: true positive rate (correct trials correctly classified). TNR: true negative rate (incorrect trials correctly classified).*

To characterise which channels drove classification, we performed an ablation analysis: each channel's contribution to the spatial covariance matrix was removed in turn, and the resulting drop in leave-one-subject-out accuracy was taken as that channel's importance (Fig. 7). The largest drops in accuracy occurred at occipital and parietal sites (Oz, O2, P3), with central and temporal channels (C4, T8) contributing more modestly further down the ranking. This posterior-weighted profile is consistent with a perceptual rather than motor origin for the decoded signal(Pfurtscheller and Lopes da Silva, 1999) (Fig. 6), and is a more conservative and defensible characterisation than raw classifier coefficient magnitude, which we found to be diffusely distributed across channels due to L2 regularisation and did not reliably distinguish genuinely informative channels from correlated, redundant ones.

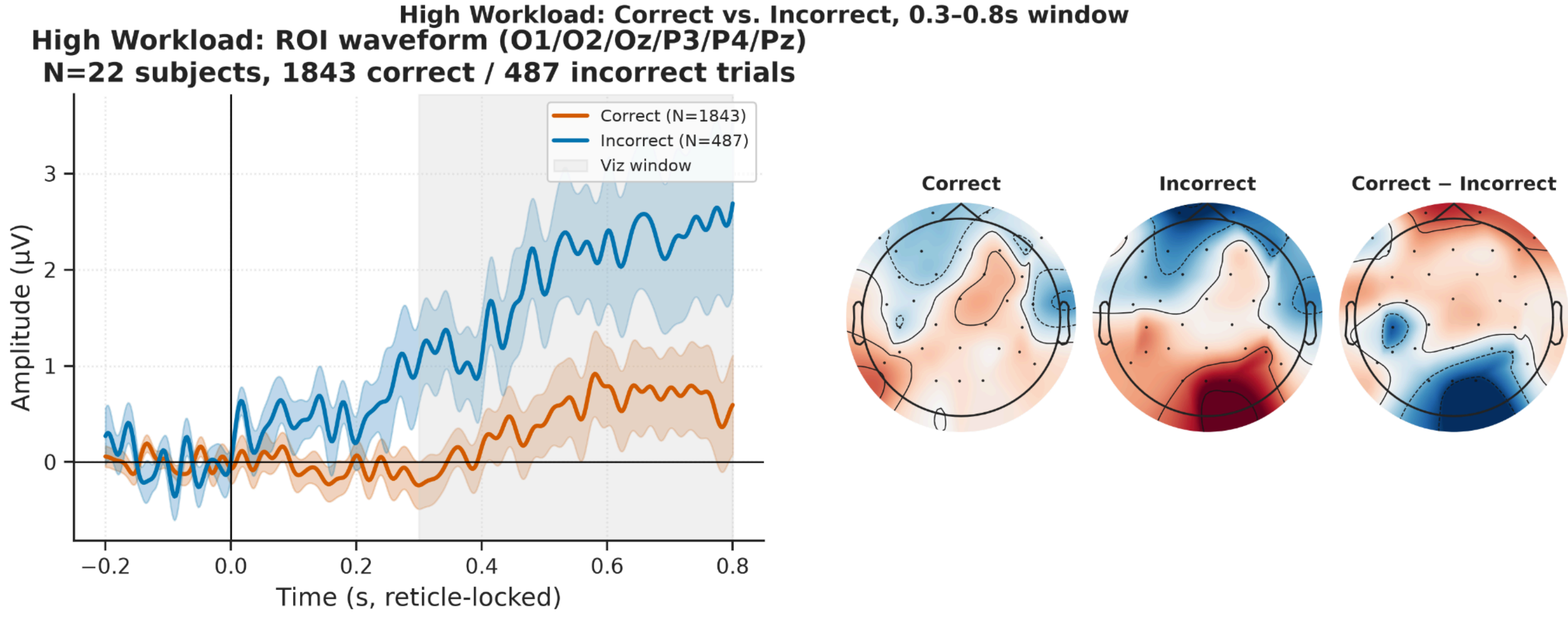

*Fig. 6 . Grand-average Correct versus Incorrect waveform and scalp topography, High Workload, fixed 0.3-0.8 s visualisation window. Left: mean amplitude ± SEM at occipito-parietal sites (O1, O2, Oz, P3, P4, Pz). Right: scalp topography for Correct, Incorrect, and their difference. Low Workload is not shown, as its classifier is not mechanistically interpreted (see Results).*

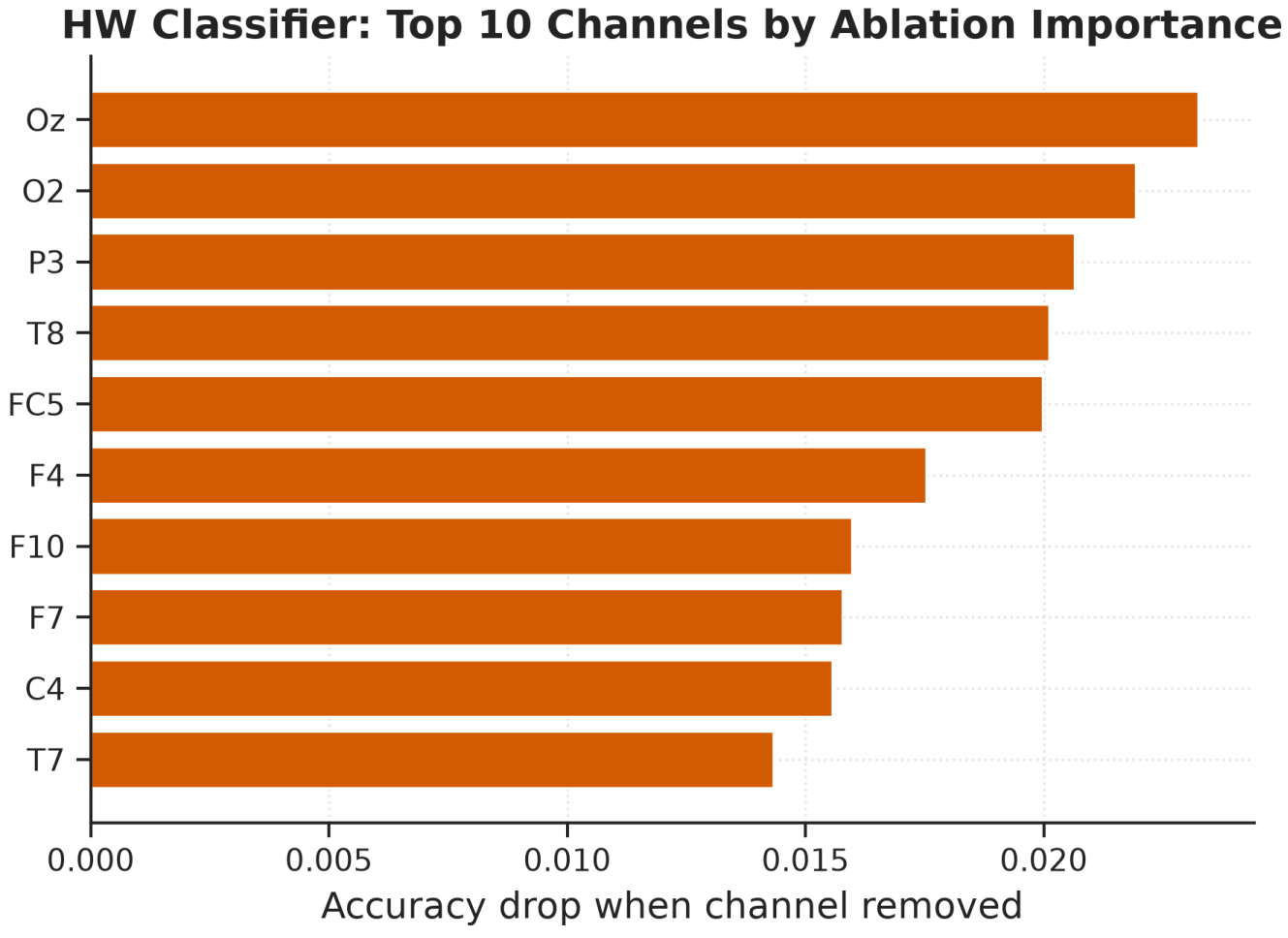

*Fig. 7 . Top ten channels ranked by ablation importance (drop in leave-one-subject-out accuracy when that channel is removed), High Workload classifier.*

**Trial Difficulty Is Explained by the Target/Non-Target Asymmetry**

To characterise what made a trial contested, we computed, for every trial with at least three available responses, a vote margin: the absolute value of the summed response directions divided by the number of respondents, ranging from 0 (perfectly split) to 1 (unanimous). Target trials showed significantly lower vote margins (more contested) than Non-Target trials, in both High Workload ($t = -5.63$, $p < .0001$, N = 147 trials) and Low Workload ($t = -4.78$, $p < .0001$, N = 144 trials; Table 4). This indicates that trial difficulty, measured independently via inter-rater disagreement, is substantially explained by the same Target-detection asymmetry established behaviourally, rather than reflecting an unrelated or unexplained source of difficulty.

| Condition | N_Trials | Mean_Margin_Target | Mean_Margin_NonTarget | t | p |
|---|---|---|---|---|---|
| High Workload | 147 | 0.48 | 0.69 | -5.63 | 9.05E-08 |
| Low Workload | 144 | 0.71 | 0.88 | -4.78 | 4.29E-06 |

*Table 4. Vote margin by Target/Non-Target status. Lower margin indicates a more contested trial (closer to an evenly split vote).*

**Pre-emptive Neural Weighting Improves Team Accuracy on Contested Trials Under High Workload**

We next asked whether this individual-level signal, used pre-emptively to weight each team member's vote before the team's collective decision, provided value at the team level. Because majority vote alone already performs strongly on the full trial set, rising to 94-99% accuracy by team size 16 in both workload conditions (Fig. 8), any weighting scheme can only plausibly change a team's outcome on trials where the unweighted vote is tied. We therefore restricted this analysis to true-tie trials, using bootstrap-resampled simulated teams (500 draws per team size and session) at sizes of 2, 4, 8, and 16.

On these contested trials under High Workload, weighting each vote by the classifier's pre-emptive probability output substantially improved team accuracy, rising from 57.2% at team size 2 to 88.2% at team size 16, while unweighted vote and average individual accuracy remained at approximately 50%, the base rate implied by a tied vote (Fig. 9). Under Low Workload, the same weighting scheme produced the opposite effect, collapsing team accuracy toward 10% by team size 8 (Fig. 9), consistent with the unreliable, low-true-negative-rate classifier established above: on the one trial subset where its weighting could influence the outcome, it was confidently wrong rather than merely uninformative.

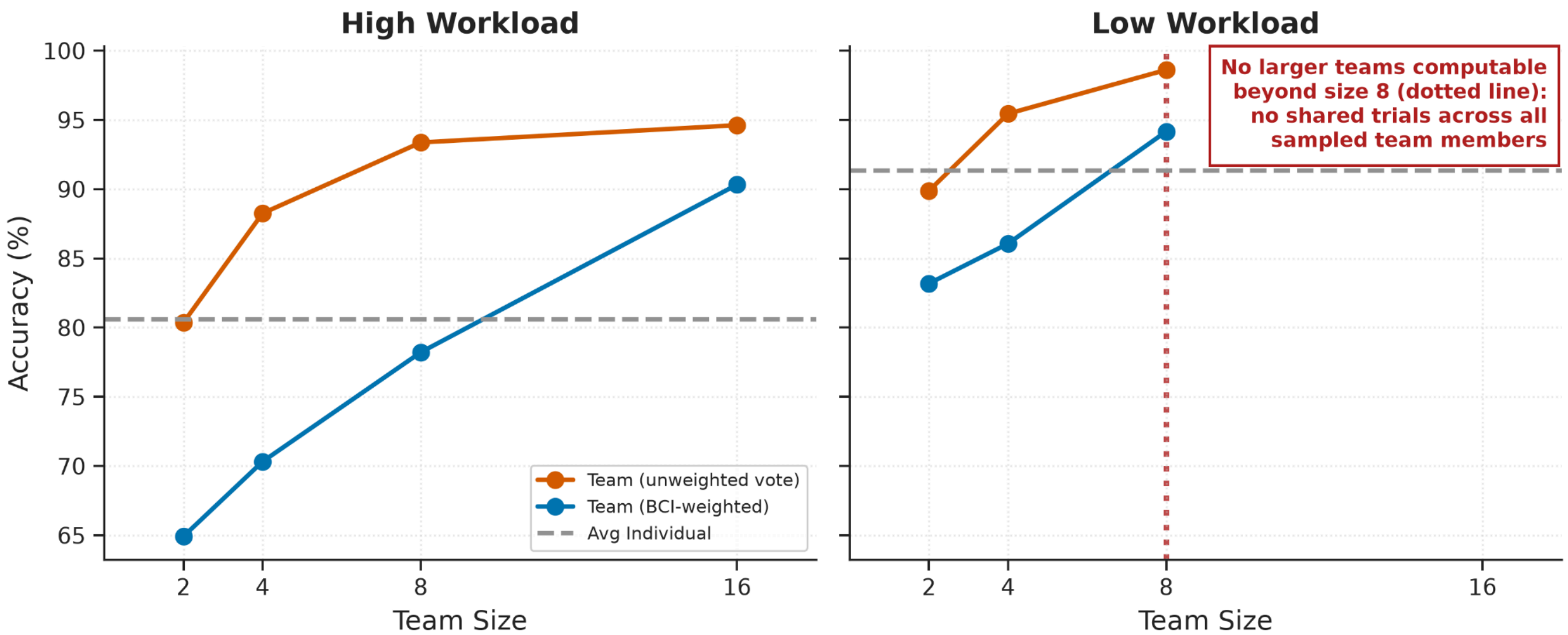

*Fig. 8. Team accuracy across all trials (no tie restriction), unweighted majority vote versus neural-weighted vote, by team size and workload. Majority vote alone already scales close to ceiling, motivating the tie-restricted analysis in Fig. 9.*

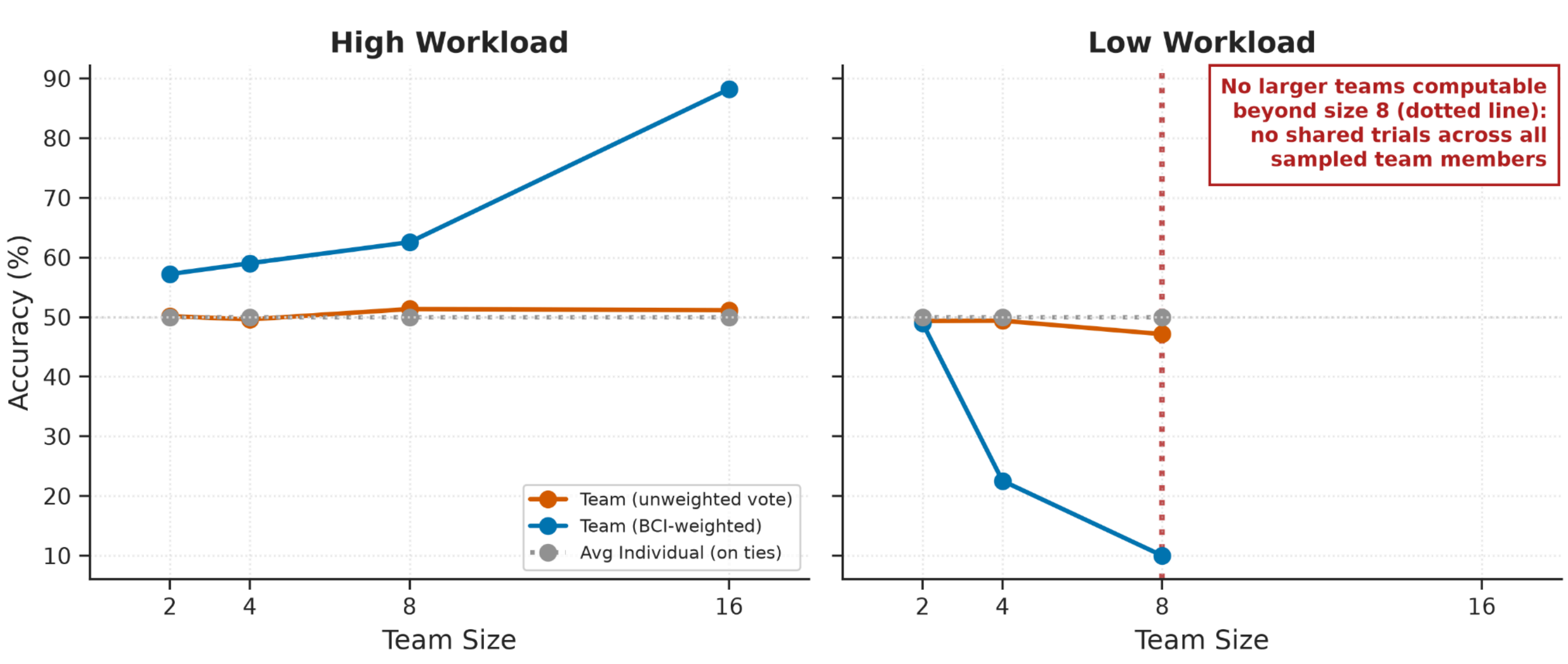

*Fig. 9. Team accuracy on true-tie trials (trials on which the unweighted vote was split within one vote of even), unweighted vote versus pre-emptive neural weighting, by team size and workload. Dotted line shows average individual accuracy conditioned on the same tie trials (≈50%, the base rate implied by a tie).*

**Post-Hoc Behavioural Signals Are Strong but Not Pre-emptive**

For comparison, we evaluated two post-hoc behavioural signals available only after a response is reported: stated confidence and response time (Fig. 10). Confidence-weighting was the strongest single signal tested, reaching 99.4% team accuracy at team size 16 under High Workload and 94.3% at team size 8 under Low Workload, in both cases exceeding the pre-emptive neural signal. Response-time weighting, by contrast, underperformed and declined as team size increased. This was not attributable to noise: on tie trials specifically, correct responses were significantly slower than incorrect ones (1.549s vs. 1.466s, $t = 7.43$, $p < .0001$), the reverse of the overall speed-accuracy relationship observed across all trials(Heitz, 2014) (faster responses were more accurate overall, $p = .0045$). Because response-time weighting assumes faster responses are more trustworthy throughout, it is systematically miscalibrated specifically on the trials it is used to adjudicate. Combining the pre-emptive neural signal with response time did not improve on the neural signal alone in either workload condition (Fig. 10).

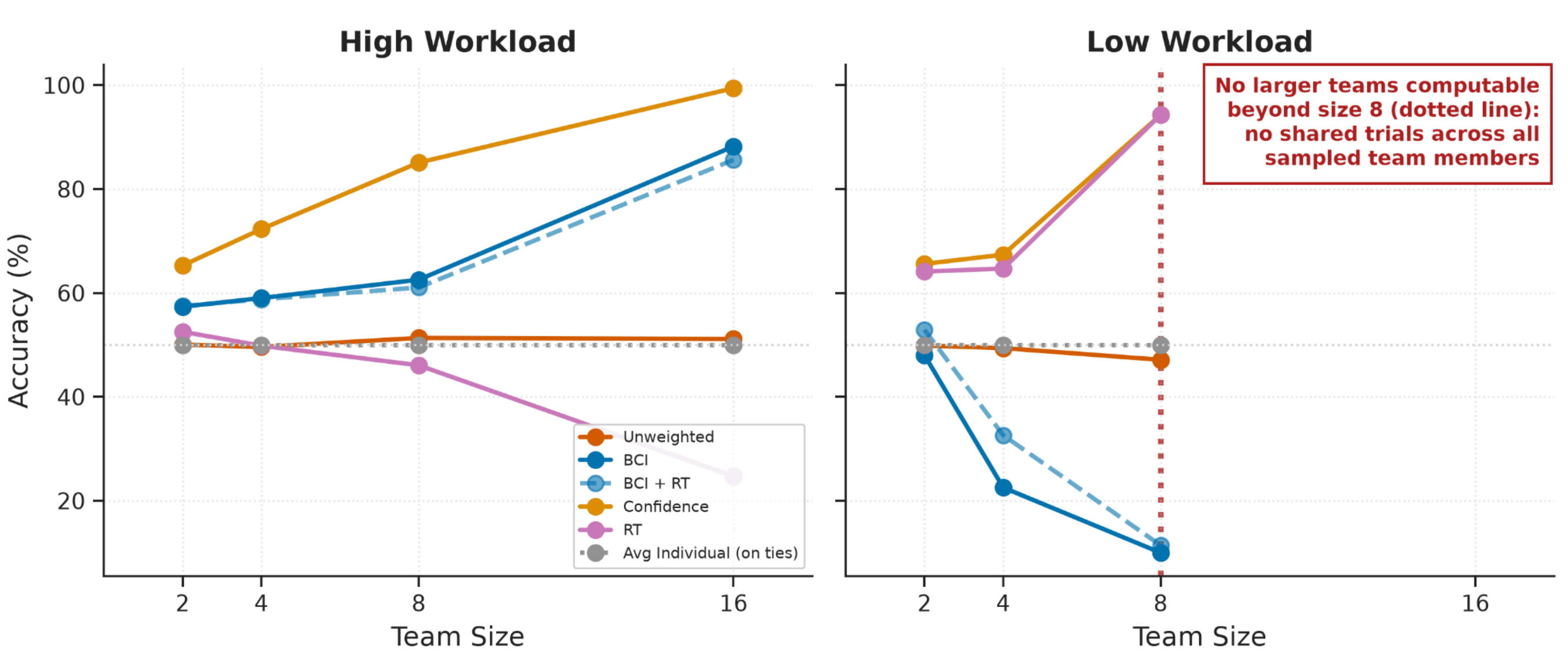

*Fig. 10. Team accuracy on true-tie trials across five aggregation schemes: unweighted vote, pre-emptive neural weighting (BCI), BCI combined with response time, post-hoc confidence weighting, and post-hoc response-time weighting alone, by team size and workload.*

## Discussion

The present study set out to test whether spatial-covariance EEG features could provide a genuinely pre-emptive signal of an operator's decision correctness, available before a response is committed, and whether such a signal depends on cognitive workload. We found that it can, but only under High Workload: a Riemannian tangent-space classifier, trained on a strictly motor-safe window, decoded Correct from Incorrect decisions with above-chance, statistically robust accuracy under High Workload, while the equivalent classifier under Low Workload was unreliable, a consequence of near-ceiling behavioural performance leaving too few genuine errors to learn from. At the team level, this signal, used pre-emptively, produced substantial accuracy gains on contested trials specifically under High Workload, while the same weighting scheme actively harmed team accuracy under Low Workload. Together, these findings indicate that a neural decision-reliability signal is not a general-purpose team augmentation tool, but one whose value is conditional on the cognitive state it is deployed under.

### The Value of a Pre-emptive Signal

The distinction between pre-emptive and post-hoc evidence, introduced in the Introduction, is not merely conceptual. Post-hoc confidence weighting outperformed the neural signal at the team level in absolute terms, reaching near-ceiling accuracy on contested trials in both workload conditions. This is an important result in its own right: it demonstrates that operators retain meaningful metacognitive access to their own decision reliability even under high workload, consistent with prior work on the dissociability of sensory evidence and behavioural report (Fleming and Lau, 2014; Wokke et al., 2020). However, by definition, confidence cannot inform a decision that has not yet been made, and a team aggregation mechanism that must wait for every member to respond and self-report before weighting their input cannot intervene in time-critical or continuously evolving decisions. The neural signal's comparatively modest team-level gains under High Workload should therefore be read not as a weaker result than confidence weighting, but as a different and complementary one: the only signal tested here that is, in principle, actionable before a response is finalised. Operational systems requiring real-time intervention, rather than post-decision review, have no substitute for this class of signal, regardless of how well post-hoc measures perform once available.

## Mechanistic Origin and Its Implications

Ablation analysis indicated that the decoded signal was carried predominantly by posterior occipito-parietal channels (Oz, O2, P3), rather than by central or temporal sites more classically associated with motor preparation(Pfurtscheller and Lopes da Silva, 1999). This is a meaningful methodological safeguard: because features were extracted from a strictly motor-safe window ending 100 ms before each trial's own response, and because the channels driving classification are posterior rather than central, the decoded signal is unlikely to reflect residual motor preparation leaking into the analysis window. Instead, this profile is consistent with the classifier detecting a perceptual or evidence-accumulation process, plausibly related to the neural correlates of error monitoring and metacognitive evaluation reported prior to, or independent of, explicit awareness (Di Gregorio et al., 2020; Schnuerch et al., 2024; Yeung and Summerfield, 2012). We note, however, that this is a correlational, channel-level characterisation rather than a source-localised one, and we return to this as a limitation below.

## Trial Difficulty and the Boundary Conditions of the Signal

The finding that contested trials, those on which teams could not agree, were substantially explained by the same Target-detection asymmetry established behaviourally (Target trials being intrinsically harder to detect than Non-Target trials to reject) offers a coherent account of when this signal is likely to matter operationally. Because team-level weighting can only influence outcomes on trials where members already disagree, and because such disagreement is concentrated on a specific, identifiable class of trial, the practical value of a pre-emptive neural signal is not diffuse across all decisions but concentrated where teams most need external adjudication. This is a more precise and operationally useful characterisation than a general claim that neural signals "improve team decision-making", and offers a template for identifying, in other domains, which specific decisions a cBCI system is likely to add value to before deployment.

## Limitations

Several limitations warrant caution in interpreting and generalising these findings. First, the Low Workload classifier's unreliability, while explained and consistent across independent analyses (raw accuracy, true negative rate, ablation-eligible sample size, and team-level collapse), is itself a consequence of small error-trial counts in that condition; a task producing a more even balance of correct and incorrect trials under low workload would be needed to establish whether the absence of a decodable signal under low workload reflects a genuine absence of the underlying neural process, or simply insufficient data to detect it. Second, response-time weighting's failure was traced to a specific, workload- and trial-difficulty-dependent reversal of the speed-accuracy relationship; while this explanation is well-supported by the data reported here, it may not generalise to tasks with different response dynamics, and should not be read as a general claim that response time is an uninformative signal in other contexts. Third, team-level results are derived from bootstrap-resampled simulated teams drawn from a single-task, single-session dataset per participant, rather than from teams who trained or worked together; whether these gains persist in genuinely interactive team settings, where members can observe and adapt to one another, remains an open question. Finally, the task, while implemented in an ecologically motivated continuous virtual reality environment, remains a single perceptual discrimination paradigm; generalisation to other decision domains (e.g. diagnostic, financial) would require independent validation.

## Conclusion

EEG-based decoding of decision correctness, using a Riemannian spatial-covariance approach applied to a strictly motor-safe window, provides a genuinely pre-emptive signal of decision reliability that is real, mechanistically grounded in posterior perceptual channels, and translatable into meaningful team-level accuracy gains, but only under high cognitive workload. Under low workload, where behavioural performance is already near ceiling, the same signal and the same aggregation mechanism are unreliable and actively detrimental. This workload-conditionality is not a limitation to be explained away but the central, actionable finding of the study: collaborative brain-computer interface systems should be deployed selectively, informed by the cognitive state of the operators they are intended to support, rather than applied uniformly across all conditions.

## Data Availability

The full dataset supporting this study, comprising raw and preprocessed EEG recordings, behavioural logs, and analysis code, exceeds 20 GB and is not practicable to host on standard manuscript-linked repositories (e.g. Figshare, OSF) given prevailing file-size limits. The complete dataset is maintained in a version-controlled GitLab repository and is available from the corresponding author upon reasonable request. Access will be granted to qualified researchers for the purposes of verification, reproduction, or extension of the reported analyses, subject to any applicable data protection and ethical approval requirements arising from the study's MoDREC approval (Reference: RQ0000037929).

## Code Availability

The custom Python code developed for this study, including the motor-safe spatial covariance feature extraction pipeline, Riemannian tangent-space classification, and team aggregation simulation, is implemented as a series of Marimo notebooks spanning all three studies in this research programme. These notebooks are hosted in a sister GitLab repository and are available on reasonable request, subject to approval by the Defence Science and Technology Laboratory (Dstl). The specific Marimo notebooks underlying the analyses reported in this manuscript will additionally be made available via the journal's designated open-source code repository at the time of publication.

## Acknowledgements

This research was supported by the Defence Science and Technology Laboratory (Dstl) under grant reference [MoDREC, Reference: RQ0000037929]. The authors wish to thank the participants for their time and engagement with the experimental protocol.

## Author Contributions

S.F. supervised the project, acquired funding. C.B. conceptualised the study, engineered the machine learning architecture, analysed the data, and wrote the original manuscript draft. S.H. conducted the virtual reality experiments, managed the electroencephalographic data acquisition, and contributed to the methodology. A.N., R.P., and C.C. provided critical revisions to the analytical pipeline and the theoretical framework. T.R. provided domain expertise for remote operations and contributed to the experimental design. All authors reviewed and approved the submitted version.

## Competing Interests

The authors declare no competing interests.